# Evidence for equilibrium excitons and exciton condensation in monolayer WTe$_2$

Bosong Sun[1], Wenjin Zhao[1], Tauno Palomaki[1], Zaiyao Fei[1], Elliott Runburg[1], Paul Malinowski[1], Xiong Huang[2,3], John Cenker[1], Yong-Tao Cui[2], Jiun-Haw Chu[1], Xiaodong Xu[1], S. Samaneh Ataei[4,5], Daniele Varsano[4], Maurizia Palummo[6], Elisa Molinari[4,7], Massimo Rontani[4*], David H. Cobden[1*]

[1]Department of Physics, University of Washington, Seattle, WA 98195, USA
[2]Department of Physics and Astronomy, University of California, Riverside, CA 92521, USA
[3]Department of Materials Science and Engineering, University of California, Riverside, CA 92521, USA
[4]CNR-NANO, Via Campi 213a, 41125 Modena, Italy
[5]Dept of Physics, Shahid Beheshti University, Evin, Tehran 1983969411, Iran
[6]INFN, Dept of Physics, University of Rome Tor Vergata, Via della Ricerca Scientifica 1, 00133 Roma, Italy
[7]Dept of Physics, Informatics and Mathematics, University of Modena and Reggio Emilia, Via Campi 213a, 41125 Modena, Italy
Email: massimo.rontani@nano.cnr.it; cobden@uw.edu

A single monolayer of the layered semimetal WTe$_2$ behaves as a two-dimensional topological insulator, with helical conducting edge modes surrounding a bulk state that becomes insulating at low temperatures. Here we present evidence that the bulk state has a very unusual nature, containing electrons and holes bound by Coulomb attraction—excitons—that spontaneously form in thermal equilibrium. On cooling from room temperature to 100 K the conductivity develops a V-shaped dependence on electrostatic doping, while the chemical potential develops a ~43 meV step at the neutral point. These features are much sharper than is possible in an independent-electron picture, but they can be largely accounted for by positing that some of the electrons and holes are paired in equilibrium. Our calculations from first principles show that the exciton binding energy is larger than 100 meV and the radius as small as 4 nm, explaining their formation at high temperature and doping levels. Below 100 K more strongly insulating behavior is seen, suggesting that a charge-ordered state forms. The observed absence of charge density waves in this state appears surprising within an excitonic insulator picture, but we show that it can be explained by the symmetries of the exciton wave function. Monolayer WTe$_2$ therefore presents an exceptional combination of topological properties and strong correlations over a wide temperature range.

In a monolayer semimetal, low carrier densities combined with reduced dimensionality provide the conditions for strong correlation effects. One possible form of such correlations is pairing of electrons and holes in the equilibrium state to form excitons. At low temperatures, such excitons could condense to form an excitonic insulator[1-4]. However, exciton formation is expected to be easily disrupted by free carriers which screen the binding interaction, and thus to occur only at low temperatures and near charge neutrality. A number of materials have been mooted as excitonic insulator candidates[5-10], but there is no consensus as to whether any of them truly contains excitons in equilibrium, either as an incoherent gas or as a coherent condensate, except in the case of bilayer heterostructures at high magnetic fields[11-14].

WTe$_2$ is a layered semimetal, but an exfoliated WTe$_2$ monolayer behaves[15-17] as a two-dimensional topological insulator, exhibiting helical conducting edge modes, which becomes superconducting when electrostatically doped[18,19]. The monolayer has the 1T' structure shown in Fig. 1a. Its bands are spin degenerate due to inversion symmetry, and near the Fermi energy, $E_F$, there is a valence ($v$) band maximum at $\Gamma$ flanked by two conduction ($c$) band minima located at $k_x = \pm k_\Lambda$ as sketched in Fig. 1b. Some tunneling spectroscopy measurements[20], angle-resolved photoemission[20,21], and density functional theory (DFT) calculations[20-23] point to a positive band gap, $E_g$, of the order of 50 meV, while others suggest overlapping bands[15,24]. Noting that in the photoemission spectra the $v$ and $c$ band photoemission features are broad enough that they overlap at least somewhat, we use thick lines in the sketch to signify this uncertainty in $E_g$.

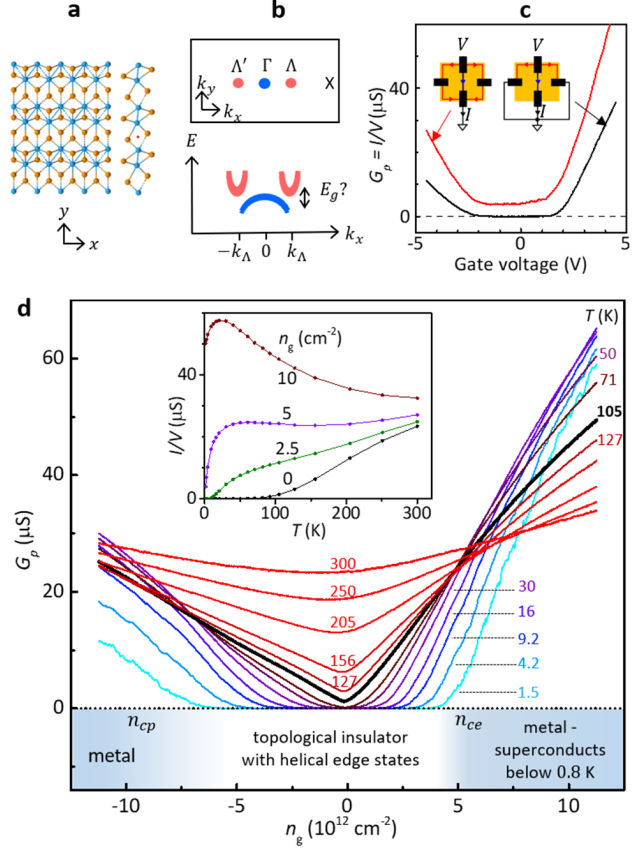

**Figure 1. Bulk conduction measurements on monolayer WTe$_2$.** (a) 1T' structure of monolayer WTe$_2$. The $x$-axis is taken to be along the zigzag W chains. (b) Schematic Brillouin zone (above) and bands near $E_F$ (below). (c) Technique used to exclude edge conduction: when the side contacts are grounded the measured current between top and bottom contacts must flow through the interior and the edge conduction plateau disappears. ($G_p \equiv I/V$). These measurements are on device MW2 at $T$ =10 K. (d) $G_p$ vs gate-induced density $n_g$ at a series of temperatures $T$ on the same device. Inset: Temperature dependence at positive values of $n_g$. Below: regimes of insulating, metallic and superconducting behavior identified in previous work.

This band structure immediately invokes the possibility that excitons could occur in equilibrium in monolayer WTe$_2$, and therefore that the insulating state might not be a simple band insulator[25-28]. In this work, we argue that the behavior of the conductivity and the electron chemical potential, even well above 100 K, is impossible to reconcile with an independent particle picture, and strongly indicates the presence of excitons in the equilibrium state. Our first-principles calculations of exciton dispersion and Bohr radius support this conclusion. The insulating behavior below 100 K suggests a charge-ordered state, but in an excitonic insulator one would normally expect charge density waves, no signs of which are seen in scanning tunneling microscopy or Raman spectroscopy. To explain this, we show that the entanglement of spin, orbital, and valley degrees of freedom hides the charge order, as the contributions to the density wave paired through time reversal cancel out.

The measurements were made on exfoliated monolayer WTe$_2$ flakes with platinum contacts, encapsulated by hBN, with graphite gates either below or above (see Supplementary Information SI-1). To study the sheet conductivity while excluding edge conduction we used the approach illustrated in Fig. 1c. As indicated in the insets, a bias $V$ is applied to one contact and the current $I$ flowing to ground through an opposite contact is measured. When the intervening side contacts are grounded this current must flow through the bulk. The edge current, which produces the plateau at low $V_g$, is thereby eliminated and the "partial conductance" $G_p \equiv I/V$ reflects the sheet conductivity, $\sigma$, via $G_p^{-1} \approx \beta/\sigma + R_c$, where $\beta$ is a geometrical factor considerably larger than one and $R_c$ is contact resistance. The gate-induced areal number density $n_g$ is deduced from the voltages applied to the graphite gate(s) and the geometric capacitances.

Figure 1d shows measurements of $G_p$ vs $n_g$ and temperature $T$. These characteristics are not measurably affected by either a normal displacement field or a normal magnetic field of 14 T, confirming the rejection of edge conduction which is highly sensitive to magnetic field at low temperatures[29]. (A recent paper[30] reports surprising quantum oscillations at low $n_g$, but we have not seen these in any device). On cooling from room temperature to 100 K, $G_p$ vs $n_p$ develops a sharp "V" shape centered close to $n_g = 0$. We have seen consistent behavior across a dozen devices, though the sharpness of the V varies, probably as a result of variable sample homogeneity. As shown in the inset, for positive $n_g$ smaller than a value $n_{ce} \approx +5 \times 10^{12}$ cm$^{-2}$, $G_p$ decreases monotonically on cooling, whereas for $n_g > n_{ce}$ it initially increases. For negative $n_g$ (hole doping), a similar but less clear-cut transition occurs around $n_{cp} \approx -10 \times 10^{12}$ cm$^{-2}$. As indicated in a band below the figure, these values of $n_{ce}$ and $n_{cp}$ are consistent with the thresholds for metallic behavior reported in previous work where it was found that the metallic state at $n_g > n_{ce}$ becomes superconducting below about 0.8 K.

Below 100 K, as $T$ decreases $G_p$ collapses over an increasingly wide range of $n_g$. This insulating behavior allows the edge conduction to dominate in normal geometries. We used microwave impedance microscopy[31] (MIM – Supplementary Information SI-2) on devices with no top gate to confirm that this is not a contact effect, as well as to detect any cracks in the monolayer WTe$_2$ which could invalidate the measurements. Figure 2a is a MIM image of device MW10 at 11 K. Red dashed lines mark the edges of the monolayer WTe$_2$ flake, and the wiggly bright lines are cracks. Fig. 2b shows the MIM-derived conductivity, $\sigma_{MIM}$, measured in the center of the white dashed square. Like $G_p$, it collapses over a range of $n_g$ that grows as $T$ falls, indicated by the dotted white contour which is drawn at $\sigma_{MIM} \approx 0.1$ μS. The drop-off of $G_p$ at low temperatures even at large $n_g$, evident in Fig. 1d, can be explained by the fact that the monolayer adjacent to the metal contacts is partially screened from the gates and so is less doped and remains insulating. In addition, the contact pattern in MW10 was aligned with the crystal axes, as determined by Raman spectroscopy (Fig. 2a inset; Supplementary Information SI-3), allowing us to compare conductivities along the *x*- and *y*-axes (Figs 2c and d; see also Supplementary Information SI-4). We see that there is substantial gate-dependent anisotropy, with the lowest conductivity occurring for p-doping parallel to the *x*-axis. This is consistent with the direction in which the valence band edge has a large effective mass.

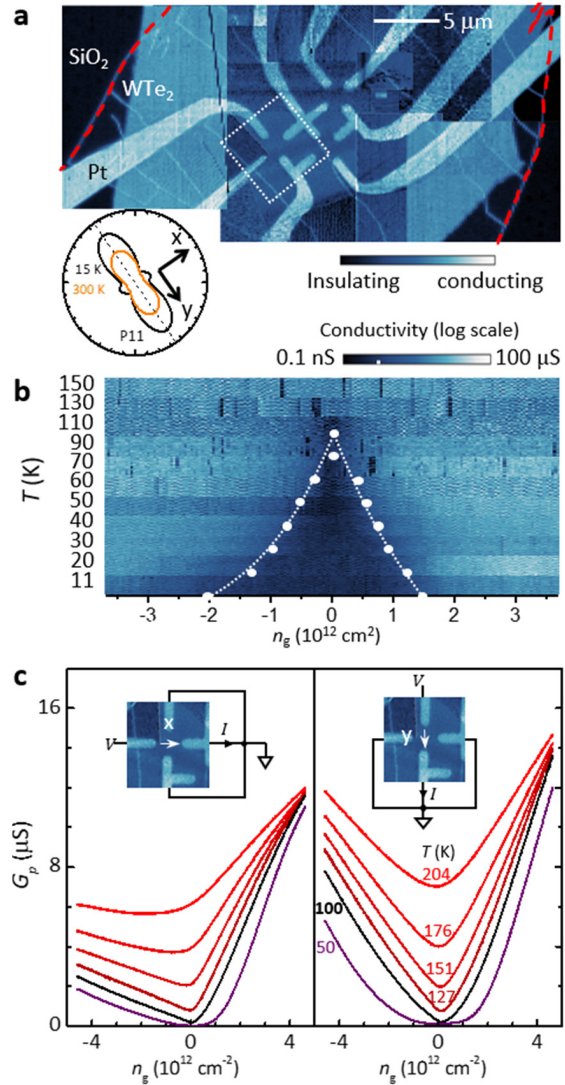

**Figure 2. Local conductivity imaging and anisotropy.** (a) Microwave impedance microscopy (MIM) image of device MW10. Here the uncalibrated imaginary MIM signal is plotted for the purpose of identifying conducting features. Multiple conducting cracks are seen running from upper left to lower right. Inset: polar plot of the Raman peak P11 (210 cm$^{-1}$) intensity, used to determine the crystal axes. (b) Conductivity vs. $n_g$ and $T$ at 2 GHz deduced from MIM measurements in the center of the white dotted square in (a). (c) Partial conductance parallel to the *x*-axis (left panel) and *y*-axis (right), measured in the white dotted square using the configurations shown in the insets.

To measure the chemical potential, we study a device including a separately contacted graphene sheet in parallel with the WTe$_2$ monolayer, as indicated in Fig. 3a. Briefly, the WTe$_2$ is almost an equipotential since it has finite conductivity and carries no current. With both the graphene and bottom gate grounded, a voltage $V_g$ is applied to the top gate relative to the WTe$_2$ and the voltage $V_W$ on the WTe$_2$ is adjusted to bring the graphene conductance to a minimum. This keeps the graphene neutral and maintains zero electric field beneath the WTe$_2$. The electrostatic potential in the WTe$_2$ is thus in effect fixed to that of the graphene, so the change in $V_W$ is due to the change in chemical potential, $\Delta\mu = -e\Delta V_W$, associated with the gate-induced charge density $-en_g = \epsilon_r\epsilon_0 V_g/d$. From $V_W$ vs $V_g$ we thereby obtain $\mu(n_g)$, choosing the zero of $\mu$ at each temperature for convenience. Figure 3b shows measurements of both $\mu$ (black) and $G_p$ (red) vs $n_g$ made on device (MW12). As usual, $G_p$ forms a sharp V as a function of $n_g$ at 100 K. Meanwhile, $\mu$ exhibits a step at the center of the V which is still discernible at room temperature and which grows on cooling, saturating at ~ 40 meV in height below about 50 K. The same behavior was seen in two devices (Supplementary Information SI-5).

The variations of $G_p$ and $\mu$ with $n_g$ are impossible to reconcile with a single-particle picture. There $\mu$ and $n_g$ are related by $n_g = n_0 + \int_{-\infty}^{+\infty} D(E) f(E) dE$, where $D(E)$ is the total electron density of states, $f(E) = [1 + exp\{(E - \mu)/kT\}]^{-1}$, and $n_0$ is a constant. To match the variation of $\mu$ with $n_g$ at low temperatures, $D(E)$ must have roughly the form shown in Fig. 3c: a constant value $D_0$ in the conduction band to give a uniform slope for $n_g > 0$; a gap, $E_g$; and a $\delta$-function-like peak at the valence band edge ($E = 0$) to make $\mu$ flat for $n_g < 0$. In Fig. 3d we plot the chemical potential calculated at the same temperatures as the measurements in Fig. 3b, using this $D(E)$ with best-fit parameters $D_0 = 3.7 \times 10^{11}$ cm$^{-2}$ meV$^{-1}$ and $E_g = 43$ meV. At 150 K the step is washed out: in fact, no choice of $D(E)$ can yield a distinct step in $\mu$ whose height is less than $kT$ as is needed to match the measurements at $T \geq 150$ K. In particular, states in the gap will only smear the step more. We also plot (dotted lines) the calculated populations of electrons in the conduction band, $n$, and holes in the valence band, $p = n_g - n$, to contrast their thermally smeared dependence of $n_g$ with the sharp V-shape seen in the conductance when $T \geq 100$ K.

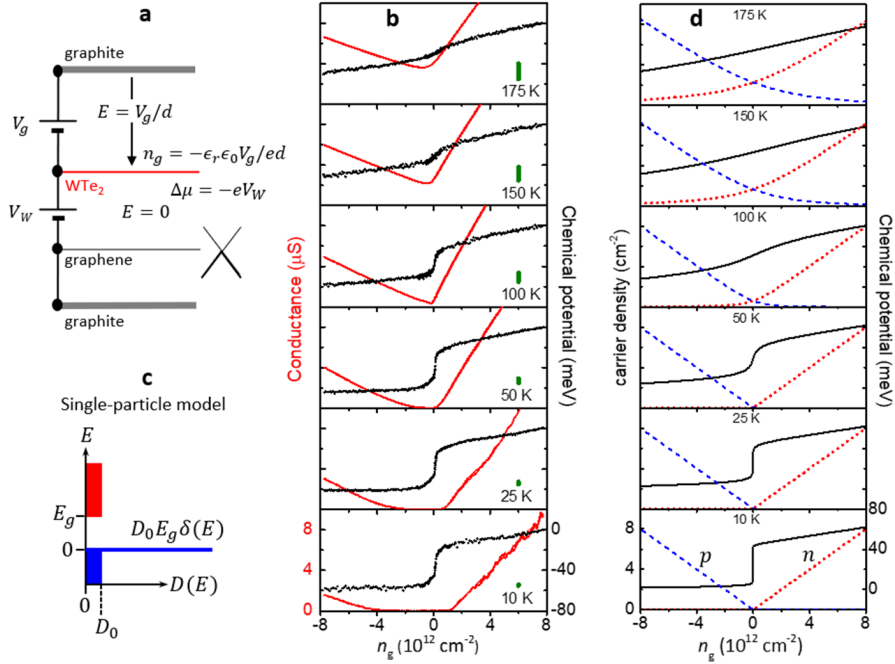

**Figure 3. Chemical potential measurements and comparison with single-particle model.** (a) Schematic of the device structure used to measure the chemical potential vs doping. In parallel with (below) the monolayer WTe$_2$ is a graphene sheet which is maintained at its Dirac point so that the electric field beneath the WTe$_2$ is zero. Doping $n_g$ and the chemical potential $\mu$ are then obtained from the voltages as shown (see text). The hBN dielectric between the layers is not shown. (b) Measurements of $\mu$ (black) and conductance $G_p$ (red) vs $n_g$ on monolayer WTe$_2$ device MW12. The length of the green bars indicates the thermal energy, $kT$. (c) Single-particle density of states $D(E)$ that reproduces the low-temperature $\mu - n_g$ behavior. (d) Calculated $\mu$ (black), electron density $n$ (red dotted line), and hole density $p$ (blue dashed line) using this $D(E)$ at the same temperatures as the measurements.

The contradictions between the single-particle picture and the observed dependence of $\mu$ and $G$ on $n_g$ and $T$ can be largely resolved simply by positing that some electrons and holes are bound as neutral excitons with density $n_x$ so that the conductivity is $\sigma = \mu_e e(n - n_x) + \mu_h e(p - n_x)$, the sum of the contributions of $n - n_x$ free electrons and $p - n_x$ free holes with respective mobilities $\mu_e$ and $\mu_h$. In Fig. 4 we compare predictions based on this equation with data from Fig. 3 (see Supplementary Information SI-6 for more details). When $n_x = 0$ we just have $\sigma = \mu_e en + \mu_h ep$, which is thermally smeared for any choice of $D(E)$. To illustrate this, in Fig. 4a we plot $p$ and $n$ calculated using the same $D(E)$ shown in Fig. 3c, at 100 K, and in Fig. 4b we plot the calculated $\sigma$ for $n_x = 0$ (blue dotted), using a mobility ratio chosen to obtain the best match to the measured conductance at 100 K (black line). However, when $n_x$ takes its maximal value, determined by the number of minority carriers $n_x = \min(n, p)$, then for $n_g > 0$ we have $\sigma = \mu_e e(n - p) = \mu_e e n_g$ while for $n_g < 0$ we have $\sigma = \mu_h e(p - n) = -\mu_h e n_g$. The result is a sharp, asymmetric "V" shape (red dashed line) which matches the measurements much better. Note that in this limit, where only the unbalanced gate-induced charge is free to move, the detailed form of $D(E)$ becomes immaterial.

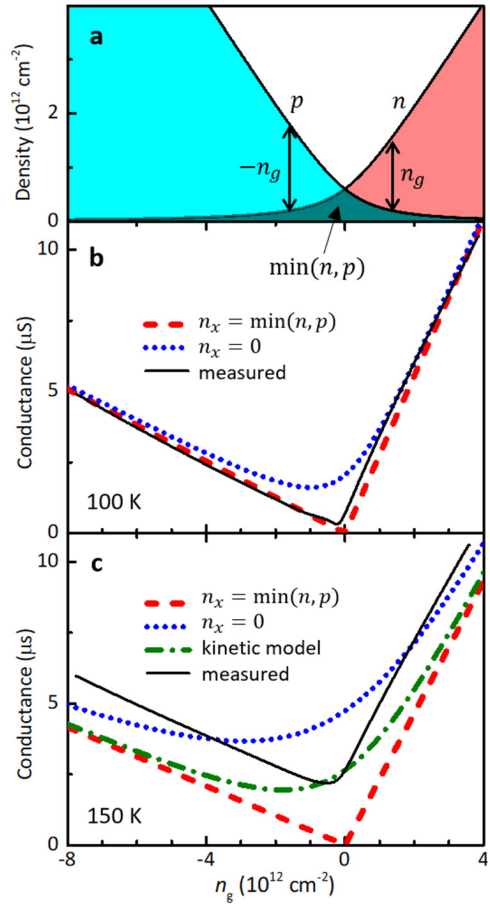

**Figure 4. Electron-hole pairing and conductivity.** (a) Total electron ($n$) and hole ($p$) densities calculated for $T = 100$ K using the single-particle density of states $D(E)$ shown in Fig. 3c. The difference is $n - p = n_g$ and the shaded area $\min(n, p)$ is the maximum possible density $n_x$ of noninteracting equilibrium excitons. (b) Measured conductance of device MW12 (black, solid line) at 100 K compared with calculated conductance for $n_x = \min(n, p)$ (red, dashed line) and for no excitons, $n_x = 0$ (blue, dotted line). (c) Same at 150 K. The additional green dash-dotted line here is obtained by allowing $n_x$ to vary in a kinetic equilibrium (see text).

As $T$ is increased from 100 K, the conductance at $n_g = 0$ rises and the sides of the "V" become shallower. This is illustrated in Fig. 4c, where we replot the conductance at 150 K (black line). The behavior remains highly incongruous with the single-particle model ($n_x = 0$, blue dotted), and is again more similar to the calculation for the case of maximal $n_x$ with slightly decreased mobilities (red dashed). However, there is now a discrepancy in the form of a vertical shift, equivalent to an extra gate-independent contribution to the conductance. The vertical shift cannot be accounted for by varying $n_x$: to illustrate this we also plot (green dash-dotted) the result of assuming that $n_x$ varies with gate voltage according to a chemical equilibrium condition, $n_x = K(n - n_x)(p - n_x)$, where $K$ is an equilibrium constant. It also cannot be reproduced by using a single-electron spectrum, for example with overlapping bands. A more sophisticated treatment of the correlated-electron system may therefore be needed to understand this aspect of the behavior.

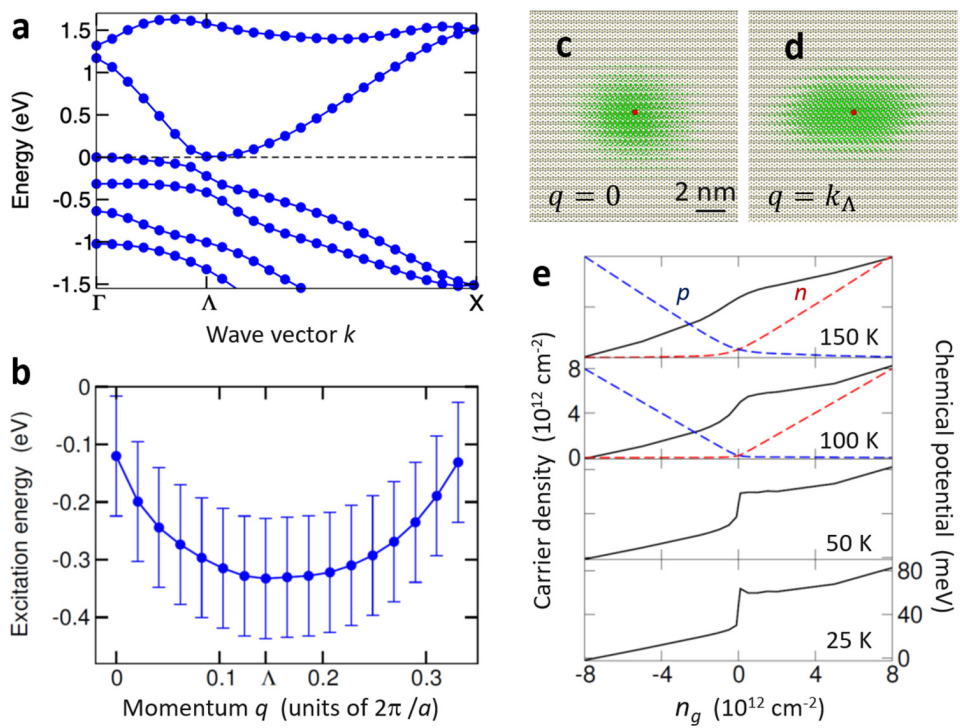

**Figure 5. Calculated exciton properties.** (a) Band structure along $\Gamma - \Lambda - X$ in Fig. 1b, obtained from first principles (DFT-PBE0 level, see Methods). Lines are guides to the eye. (b) Excitation energy of the lowest exciton vs center-of-mass momentum $\boldsymbol{q} = (q, 0)$ along the same cut, calculated by solving the Bethe-Salpeter equation. The thick curve is a guide to the eye. The excitation energy is negative over the whole range of $q$, implying instability of the ground state to the spontaneous generation of excitons. Error bars are estimated by extrapolation of Brillouin zone sampling (Methods and Supplementary Figure S7). (c-d) Wavefunction of the lowest exciton with $q = 0$ and $q = k_\Lambda$, respectively. The plot shows the conditional probability distribution of the electron for the hole located at the red dot. The plot in panel d is an average over three degenerate states. (e) Simulated behavior of $\mu$ vs $n_g$ (black solid line) for an excitonic insulator within the spinful two-band model (Methods). The density of free electrons (red dashed) and holes (blue dashed) is computed self-consistently as the population of the $c$ and $v$ bands, respectively, which are renormalized by the presence of the condensate.

Excitons that persist in equilibrium at 100 K and at doping levels above $10^{12}$ cm$^{-2}$ must have binding energy much larger than the thermal energy of ~10 meV and small size to survive screening by free charges. To see if this is plausible, we solved the exciton (Bethe-Salpeter) equation of motion from first principles, building on the DFT band structure (Fig. 5a) and including spin-orbit effects in a non-perturbative way (Methods). The resulting excitation energy vs momentum $q$ is shown in Fig. 5b. Since the binding energy only weakly depends on $E_g$ because the gap is indirect[2,10], and the value of $E_g$ is uncertain, we tuned the DFT hybrid functional to make $E_g$ vanish. The dielectric function was evaluated in the random phase approximation, and the uncertainty induced by the numerical discretization of $k$ space is shown by the error bars. The excitation energy is negative for all $q$, ranging from -100 meV for direct excitons at $q = 0$ to a minimum of -330 meV for indirect excitons made of a hole at $\Gamma$ and an electron at $\Lambda$. In Figs. 5c-d we plot the spatial profile of an exciton in the center-of-mass frame. The exciton radius is as small as 4 nm. This is comparable with the typical electron separation at the critical doping, $n_{ce}^{-1/2} = 4.5$ nm, suggesting that excitons could play a role in the insulator-metal transition at $n_{ce}$.

In interpreting the situation below 100 K in terms of an excitonic insulator, formed by condensation of the excitons, we face two problems. First, at low temperatures insulating behavior sets in over a wide doping range, approaching $n_{cp} < n_g < n_{ce}$. Conventional neutral excitonic insulator theory provides no mechanism for localizing the unbalanced charge. Nevertheless, it seems possible that the Coulomb interaction, which has a long range for these doping values, could stabilize both the excitonic phase[10] and the Wigner crystallization of unbound carriers, whose effective mass is enhanced by the opening of a many-body gap.

Second, the exciton condensate is naively expected to exhibit a charge density wave (CDW) with wave vector $k_\Lambda$, but CDWs are not seen in tunneling microscopy[24,32,33] and our detailed temperature-dependent Raman spectra show no evidence of any CDW transition (Supplementary Information SI-3). One possible explanation is that the condensate is made of direct excitons having $q = 0$, as predicted[25] for the T' phase of monolayer $MoS_2$. We have checked that this leads to no significant symmetry breaking, due to the anisotropic character of the $WTe_2$ band structure (Methods and Supplementary Figure S8). However, this state would have higher energy than a condensate made of indirect excitons with finite $q$. A more likely possibility is that the peculiar symmetries of excitons with $q = \pm k_\Lambda$ prevent the condensate from exhibiting charge order.

From the solution of Bethe-Salpeter equation we find that the lowest-energy exciton with $q = k_\Lambda$ (or $-k_\Lambda$) is three-fold degenerate and separated by 20 meV from a nondegenerate first-excited state. This energy splitting is due to the residual exchange interaction, present despite the strong spin-orbit coupling. These excitons are made of electrons and holes that populate respectively the lowest $c$ and highest $v$ band, with momentum $k$ lying near the $\Gamma\Lambda$ line. Along this high-symmetry line, both $c$ and $v$ Bloch spinors, each doubly degenerate, may be chosen as the eigenstates of the two-fold screw-axis rotation, which has complex conjugated values as irreducible representations. We label these Bloch spinors as $\zeta = \pm i$ and use them to write explicitly the exciton wave functions within a minimal two-band model, which includes both spin and orbital degrees of freedom (Methods). In addition to the rotational symmetry, we may classify the excitons according to their triplet- or singlet-like character, i.e., whether they respectively maximize or minimize the electron-hole spatial overlap. In fact, Bloch states labeled by $\zeta$ transform like spins polarized along the $x$ axis under the two-fold rotation, even if their actual spin polarization away from $\Gamma$ is zero. Specifically, the exciton wave functions that are even with respect to the screw-axis rotation are:

$$|\pm\rangle = \tfrac{1}{\sqrt{2}}\sum_{k_x,k_y} \psi^{\pm}(\boldsymbol{k})\left[c_i^{\dagger}(k_x + k_\Lambda, k_y)v_i(\boldsymbol{k}) \pm c_{-i}^{\dagger}(k_x + k_\Lambda, k_y)v_{-i}(\boldsymbol{k})\right]|0\rangle,$$

where +/− stands for singlet/triplet-like symmetry, $c_\zeta^\dagger(\mathbf{k})$ [$v_\zeta(\mathbf{k})$] creates [destroys] an electron of momentum $\mathbf{k}$ and screw-axis symmetry $\zeta$ in the $c$ [$v$] band, $|0\rangle$ is the noninteracting ground state with all $v$ states filled and $c$ states empty, and $\psi$ is the exciton wave function in reciprocal space in the center-of-mass frame, which is nodeless and even in $k_y$ (and, approximately, in $k_x$): $\psi(k_x, k_y) = \psi(k_x, -k_y)$.

In the excitonic insulator phase, the condensate is macroscopically occupied by excitons, hence the expectation value of the operator that creates an electron-hole pair, $\langle c_\zeta^+ v_{\zeta'}\rangle$, has a finite magnitude, $\phi_{\zeta\zeta'}$, and an arbitrary phase, $\theta_{\zeta\zeta'}$, i.e., the complex wave function of the condensate is $\phi_{\zeta\zeta'}\exp(i\theta_{\zeta\zeta'})$ (here $\langle\ldots\rangle$ is the average over the many-body ground state). As excitons with $\mathbf{q} = (k_\Lambda, 0)$ and $(-k_\Lambda, 0)$ are degenerate, they may separately condense, hence the wave function of the condensate has two valley components of equal magnitude, whose respective phases are related by time-reversal symmetry[10]. Furthermore, both condensate and exciton wave functions share the same screw-axis and spin-like symmetries[3], as well as the same parity in $\mathbf{k}$ space. These fundamental constraints relate all components of the condensate to a unique wave function magnitude, $\phi(\mathbf{k})$, and phase, $\theta$. If the excitonic ground state has triplet-like symmetry, one has:

$$\left\langle c_\zeta^\dagger(k_x \pm k_\Lambda, k_y)\, v_{\zeta'}(\mathbf{k}) \right\rangle = \pm(-1)[\sigma_z]_{\zeta\zeta'}\exp(\pm i\theta)\, \phi^-(\mathbf{k}).$$

For the singlet-like ground state, one replaces the $2\times 2$ Pauli matrix $\sigma_z$ with the identity matrix $\mathbf{1}$, the right hand side of the equation then reading $\mathbf{1}\exp(\pm i\theta)\,\phi^+(\mathbf{k})$. The charge/spin density wave of the excitonic insulator is dictated by the interband contribution to the expectation value of the corresponding density operator, which is proportional to the left hand side of the above equation. Therefore, we may assess the occurrence of charge order in the excitonic phase without actually computing $\phi^\pm(\mathbf{k})$, which is given by a gap equation[10] that depends on $E_g$. Indeed, $\phi$ only provides the intensity of the density modulation, whereas $\theta$ rigidly shifts the density wave with respect to the frame origin. After a lengthy but straightforward analytical calculation, the results are as follows.

There is no CDW with momentum $q = k_\Lambda$, regardless of the triplet- or singlet-like ground state. In fact, the contributions to the CDW induced by excitons that are time-reversal partners cancel out exactly, due to the entanglement of degrees of freedom: were the spin-orbit interaction negligible, these contributions would sum up instead. In brief, the charge order is hidden by the combined effect of time-reversal symmetry and spin-orbit interaction, which is ultimately related to the topological properties of WTe$_2$ (Ref. 15). The CDW may be unveiled by breaking time reversal symmetry: practically, one may split the hole states that sustain the density wave through Zeeman coupling with the magnetic field, as the hole spin is polarized along $x$ (the $\zeta$ eigenvalue reduces to the spin projection close to Γ). Finally, the ground state exhibits a spin density wave of momentum $q = k_\Lambda$, plus a weak charge modulation of period $q = 2k_\Lambda$. These features were also found in Ref. 28 by adding an intervalley scattering term to the two-band model, in the absence of which different ordered states would have been degenerate. In contrast, the hidden order we predict here is inherent to both triplet- and singlet-like symmetries of the condensate, and hence robust.

Last, we consider the behavior of $\mu$ vs $n_g$. We start by putting forward the following heuristic argument. At low temperatures ($T \lesssim 50$ K), where every minority carrier is paired, for $n_g < 0$ when an electron is added to the system it pairs with a hole, reducing the addition energy by the exciton binding energy and leading to diverging compressibility and self-consistent pinning of $\mu$ (in the charge-ordered state). For $n_g > 0$, the added electron does not pair and so $\mu$ is not affected by the binding energy; the result is a step in $\mu$ at $n_g = 0$ whose height is related to the binding energy. The decrease in the height of the step for $T \gtrsim 100$ K could be because the exciton binding weakens due to screening by the free carriers.

This scenario is supported by simulating the behavior of an excitonic insulator in the presence of free charge carriers (Fig. 5e), with both $\mu$ and the many-body gap being computed self-consistently (Methods). Here the gap has a purely excitonic origin, as the starting noninteracting phase is a semimetal. The step in $\mu$ remains clearly visible up to 100 K, in contrast with the smeared profile of the independent-electron model (Fig. 3d). This is a peculiar consequence of exciton condensation, as electrons injected for $n_g > 0$ fill in the lowest $c$ states blockading the formation of $e$-$h$ pairs—an effect due to Pauli exclusion principle and suppressed with temperature[34]. At even higher $T$ the step is smeared anyway, since the condensate is depopulated by thermal excitations and the excitonic gap melts. Whereas the simulation of Fig. 5e was done for direct excitons for the sake of illustration[25], indirect excitons will exhibit the same qualitative features.

To conclude, examination of the conductivity and thermodynamics of the monolayer topological insulator WTe$_2$ leads to the remarkable conclusion that neutral excitons are present in equilibrium, not only in an insulating charge-ordered state at low temperatures but also at temperatures above 100 K when the conductivity is substantial. The fact that this binding energy is close to the apparent band gap inferred from spectroscopy suggests that the latter might be best interpreted as a many-body gap. We note that we have seen similar characteristics in bilayer WTe$_2$ though at lower temperatures[35] (Supplementary Information SI-7) indicating that excitons with about five times weaker binding are also present in the bilayer. The low-temperature insulating many-body state of the monolayer, which isolates the helical edge modes in this 2D topological insulator, therefore probably has excitonic insulator character, and may compete with the superconducting state that becomes the ground state above a critical electron doping level.

## Methods

**Ground state calculations from first principles**

We obtained the ground-state electronic structure within density functional theory (DFT) with a plane wave basis set, as implemented in the Quantum Espresso package[36]. We fixed a kinetic energy cutoff of 80 Ry for the wave functions and used fully relativistic norm-conserving pseudopotentials[37] to include the spin-orbit interaction. We optimized the lattice parameters and atomic positions using the PBE exchange-correlation functional, the final cell parameters being $a$ = 3.52 Å and $b$ = 6.29 Å. We set the cell side along $z$ to 15 Å. We obtained the band structure using a PBE0 pseudopotential, for which we considered a small fraction of exact exchange, 2%.

**Excitation energies and exciton wave functions from first principles**

We calculated the excitation energies of excitons as well as the dispersion of the lowest-energy exciton within the framework of many-body perturbation theory[38–40], by solving the Bethe-Salpeter equation through the Yambo code[41,42] and including spin-orbit interaction in a non-perturbative way[43]. We considered the PBE0 electronic structure as a starting point and calculated the static screening in the direct term within the random phase approximation, with inclusion of local field effects; we employed the Tamm-Dancoff approximation for the Bethe-Salpeter Hamiltonian. To avoid spurious interactions among layers, we employed a truncated Coulomb cutoff technique[44]. We obtained converged excitation energies considering respectively two valence and two conduction bands in the Bethe-Salpeter matrix, the irreducible Brillouin zone being sampled up to a $48 \times 24 \times 1$ $\boldsymbol{k}$-point grid. We extrapolated the excitation energy of the lowest exciton with momentum $\boldsymbol{q} = 0$ to the limit of a dense $\boldsymbol{k}$-point grid, as shown in Supplementary Figure S7.

**Spinful two-band model**

The spinful two-band model provides the noninteracting, doubly degenerate $c$ and $v$ Bloch states of crystal momentum $\boldsymbol{k}$ that comply with the symmetry group of monolayer WTe$_2$ (the T' structure is centrosymmetric and nonsymmorphic). Within the four-dimensional spin/orbital space, the Hamiltonian is a $4 \times 4$ Hamiltonian matrix, $H_{QSH}(\boldsymbol{k})$, whose off-diagonal elements are the spin-orbit interaction terms. Here the orbital degree of freedom identifies the $c$ and $v$ Bloch states at $\Gamma$, whose energies are 'inverted' with respect to the usual order of bulk semiconductors. These two states have been variously identified in the literature as a pair of orbitals having either opposite[15,25] or like[29,45] parities under spatial inversion, leading to two different forms of the spin-orbit interaction: we label the corresponding model Hamiltonians respectively as $H_{QSH,1}(\boldsymbol{k})$ and $H_{QSH,2}(\boldsymbol{k})$. We find that the charge order is hidden in the condensate of indirect excitons regardless of the model. The reason is that the screw-axis rotation (around the W atom chain direction) maintains the same form within the spin/orbital space, entangling the degrees of freedom of its eigenstates, labelled by $\zeta = \pm i$. This is pivotal to the discussion of the main text.

In the following we detail the Hamiltonians $H_{QSH,1}(\boldsymbol{k})$ and $H_{QSH,2}(\boldsymbol{k})$. To use a notation consistent with Ref. 25, throughout this section and the next (but nowhere else) we swap the $x$ and $y$ cartesian axes with respect to the reference frame used in the main text. The first model, $H_{QSH,1}(\boldsymbol{k})$, is taken from Refs. 15 and 25 with minor adjustments. The Hamiltonian reads:

$$H_{QSH,1}(\boldsymbol{k}) = \tfrac{1}{2}[\epsilon_u(\boldsymbol{k}) + \epsilon_g(\boldsymbol{k})]1_\tau \otimes 1_\sigma + \tfrac{1}{2}[\epsilon_u(\boldsymbol{k}) - \epsilon_g(\boldsymbol{k})]\tau_z \otimes 1_\sigma + \hbar v_2 k_y \tau_x \otimes \sigma_x \, ,$$

where $v_2$ is the spin-orbit coupling parameter, $\tau_x, \tau_y, \tau_z$ and $\sigma_x, \sigma_y, \sigma_z$ are $2 \times 2$ Pauli matrices in orbital and spin space, respectively, and the $2 \times 2$ unit matrices are $1_\tau$ and $1_\sigma$. The diagonal matrix elements, $\epsilon_l(\boldsymbol{k})$ with $l = u, g$, are the band energies in the absence of spin-orbit interaction, which are inverted at $\Gamma$ and cross at the points $\pm k_\Lambda$ of the $\Gamma Y$ line in the Brillouin zone. The energies $\epsilon_l(k_x, k_y)$ are even with respect to both $k_x$ and $k_y$ axes. The corresponding Bloch states transform like $p_y$ and $d_{yz}$ orbitals at $\Gamma$. The functional dependence of $\epsilon_l$ on $\boldsymbol{k}$ differs from the effective-mass expression given in Ref. 25, but its precise form is irrelevant to the discussion of the main text, solely based on symmetry arguments. We discard the off-diagonal spin-orbit term linear in $\sigma_y$ considered in Ref. 25, the total Hamiltonian being now even in $k_x$, $H_{QSH,1}(k_x, k_y) = H_{QSH,1}(-k_x, k_y)$. This choice agrees with the evidence that the spin-orbit field lies in the $xz$ plane[29]. Furthermore, we find that the spin-orbit term proportional to $\sigma_x$ provides a good matching with the strongly anisotropic DFT bands, as we checked through comparison with our own first-principles calculations. In the spin/orbital space, the inversion operator reads $I = -\tau_z \otimes 1_\sigma$ and the screw-axis rotation around the $y$ axis is $C_{2y} = i\tau_z \otimes \sigma_y \exp(ik_y a/2)$, with $a$ being the lattice constant in the direction parallel to the W chains.

The second model, $H_{QSH,2}(\boldsymbol{k})$, is taken from Ref. 29 and builds on the four-band tight-binding Hamiltonian proposed in Ref. 45 to improve the matching between model and DFT bands. The $c$ and $v$ orbital states are Wannier functions, respectively an antibonding combination of $d_{x^2-y^2}$-type orbitals localized on W atoms [energy $\epsilon_W(\boldsymbol{k})$] and a bonding superposition of $p_y$-type orbitals localized on Te atoms [energy $\epsilon_{Te}(\boldsymbol{k})$]. These Wannier functions have the same parities under inversion but opposite parities under the screw-axis two-fold rotation, like the Bloch states at Γ of our own DFT calculations. The Hamiltonian is:

$$H_{QSH,2}(\mathbf{k}) = \frac{1}{2}[\epsilon_W(\mathbf{k}) + \epsilon_{Te}(\mathbf{k})]1_\tau \otimes 1_\sigma + \frac{1}{2}[\epsilon_W(\mathbf{k}) - \epsilon_{Te}(\mathbf{k})]\tau_z \otimes 1_\sigma + \lambda_{SO}\tau_y \otimes \sigma_x,$$

where $\lambda_{SO} > 0$ is the spin-orbit coupling parameter, which is independent from $\mathbf{k}$. For the sake of simplicity, here we have neglected the spin-orbit term proportional to $\sigma_z$ proposed by Ref. 29. Within the envelope function approximation, the band energies $\epsilon_W(\mathbf{k})$ and $\epsilon_{Te}(\mathbf{k})$ are provided by the tight-binding calculation of Ref. 45 and are even with respect to both $k_x$ and $k_y$ axes. The inversion operator now reads $\mathbf{I} = 1_\tau \otimes 1_\sigma$ whereas the screw-axis rotation around the $x$ axis is again $\mathbf{C}_{2y} = i\tau_z \otimes \sigma_y \exp(ik_y a/2)$.

**Eigenstates of the screw-axis rotation**

The eigenvectors of $H_{QSH,1}(\mathbf{k})$ that were explicitly given in Ref. 25 are spin polarized along the direction perpendicular to the W atom chains. Here we use the notation $|\mathbf{k}, \lambda\rangle$ to identify these eigenvectors, which belong to either the $c$ or $v$ band and have spin polarization $\lambda = \uparrow, \downarrow$. In the main text we introduced an alternative, equally legitimate set of eigenvectors, which are simultaneous eigenstates of $H_{QSH,1}(\mathbf{k})$ and $\mathbf{C}_{2y}$ (the latter with eigenvalues $\zeta = \pm i$), by applying a unitary rotation of the basis for any wave vector $\mathbf{k}$. Explicitly, $|\mathbf{k}, \zeta = -i\rangle = (i|\mathbf{k}, \uparrow\rangle + |\mathbf{k}, \downarrow\rangle)/\sqrt{2}$, $|\mathbf{k}, \zeta = +i\rangle = (-i|\mathbf{k}, \uparrow\rangle + |\mathbf{k}, \downarrow\rangle)/\sqrt{2}$, with $\mathbf{C}_{2y}|\mathbf{k}, \zeta = -i\rangle = -i\exp(ik_y a/2)|\mathbf{k}, \zeta = -i\rangle$ and $\mathbf{C}_{2y}|\mathbf{k}, \zeta = +i\rangle = i\exp(ik_y a/2)|\mathbf{k}, \zeta = +i\rangle$, as may be checked by direct substitution. Importantly, the states $\zeta = \pm i$ are not spin polarized (except at $\mathbf{k} = 0$), since the spin and orbital degrees of freedom are now entangled. Note that $|\mathbf{k}, \zeta = -i\rangle$ and $|-\mathbf{k}, \zeta = +i\rangle$ are time-reversal mates, with $\Theta|\mathbf{k}, \zeta = -i\rangle = i|-\mathbf{k}, \zeta = +i\rangle$, the time-reversal operator being $\Theta = i1_\tau \otimes \sigma_y K$ ($K$ is the complex conjugation operator).

The simultaneous eigenvectors of $H_{QSH,1}(\mathbf{k})$ and $\mathbf{C}_{2y}$ that we use to build the excitonic insulator ground state (within the envelope function approximation) are as follows. The $c$ band state with $\mathbf{k} = (0, k_\Lambda)$ and $\zeta = +i$ is $(-1+i)/(2\sqrt{2})[-i, i, 1, 1]$; the $v$ band state with $\mathbf{k} = 0$ and $\zeta = +i$ is (approximately) $(1/\sqrt{2})[0, i, 0, 1]$. The states with $\zeta = -i$ as well as those with $\mathbf{k} = (0, -k_\Lambda)$ are obtained through time reversal and inversion transformations. Here the first and third (second and fourth) components of the four dimensional vector, $[u_{W,\uparrow}, v_{Te,\uparrow}, u_{W,\downarrow}, v_{Te,\downarrow}]$, correspond to a spinor whose orbital part is a Wannier function localized on W (Te) atoms in the crystal unit cell.

**Condensate of indirect excitons and charge/spin order**

We assess the charge (spin) order of a permanent condensate of indirect excitons of momentum $\mathbf{q} = (\pm k_\Lambda, 0)$ within the multivalley framework developed in Ref. 10. This approach, in turn, relies on the scheme to decouple the equations of motion for Green functions of Ref. 46. The theory, which deals with spinless electrons, may be straightforwardly generalized to spinors labeled by the screw-axis symmetry $\zeta = \pm i$. Indeed, for any given electron and hole species $\zeta$ and $\zeta'$, the structure of the equations of motion for Green functions that govern the condensate component $\langle c_\zeta^+ v_{\zeta'}\rangle$ remains the same, since $\zeta$ electrons pair with $\zeta'$ holes only: as far as pairing is concerned, $\zeta$ spinors behave as if they were spinless fermions. Therefore, the $\mathbf{k}$-dependent spinless Bogoliubov-Valatin-like creation operator that defines the excitonic insulator ground state in equations 20 and 21 of Ref. 10 is simply replicated for $\zeta = \pm i$, provided one specializes equation 21 to the present case of two valley components and chooses the condensate phases as shown in the main text. Finally, we compute the expectation value of the charge (spin) density operator over the excitonic ground state, after making the spin/orbital structure of $\zeta$ Bloch states (given in the previous section) explicit, the derivation being lengthy but straightforward. As discussed in the main text, in order to simply assess the occurrence of charge (spin) order without computing the density wave modulation intensity, it is not necessary to evaluate explicitly the coherence coefficients $u_k^0$ and $v_k^0$ that occur in equation 21 of Ref. 10 (these are provided by the self-consistent gap equation 3).

**Condensate of direct excitons and simulation of Fig. 5e**

We obtain the results shown in Figure 5e within the spinful two-band model $H_{QSH,1}(\boldsymbol{k})$ described above, the band energies $\epsilon_l(\boldsymbol{k})$ being parametrized through comparison with our own DFT calculations ($l = u, g$). The noninteracting ground state, in the absence of spin-orbit interaction, is taken to be a semimetal with a band overlap of 38 meV. The energy parametrization relies partly on the tight-binding model of Ref. [45] (Table III therein) and partly on ad hoc parameters. In detail, we correct the tight-binding energies by adding the terms $2(t_l' - t_l'') \cos 2k_x + 2t_l'' \cos 2|\boldsymbol{k}|$, with $t_g' = 0.149$ eV, $t_u' = 0.075$ eV (see Ref. 23), $t_g'' = 0.049$ eV, and $t_u'' = 0.055$ eV. The spin-orbit parameter is $v_2 = 10^{14}$ Å/s, and the strength of Coulomb interaction is fixed by the two-dimensional polarizability, $\alpha_{2D} = 5.5$ (notations of Ref. 25). In order to compute the chemical potential $\mu$ vs charge density $n_g$ in the many-body excitonic phase, we adapt the theory of Ref. 25, which deals with a condensate of direct excitons in an intrinsic semiconductor, to the case of a doped system, by means of a fully-self consistent calculation of both the excitonic order parameter, $\Delta_X(\boldsymbol{k})$, and $\mu$. Furthermore, we assume that the Coulomb interaction remains long-ranged for any doping value, which is supported by the evidence that charge carriers localize in a wide doping interval at low $T$. The free carriers populating the renormalized bands of the excitonic insulator are conventionally taken as noninteracting, which is the origin of the unphysical behaviour of $\mu$ for small, positive values of $n_g$ at $T = 25$ K ($d\mu/dn_g$ is negative close to the axis origin in Fig. 5e).

We have checked that the observable effects related to the breaking of inversion symmetry, due to the condensation of excitons with $\boldsymbol{q} = 0$, are negligible for WTe$_2$, contrary to the case of T'-MoS$_2$. This is related to the limited extent of the excitonic order parameter, $\Delta_X(\boldsymbol{k})$, along the Brillouin zone direction perpendicular to the ΓΛ line, which is in turn caused by the strong anisotropy of the noninteracting $c$ and $v$ bands close to Λ (compare Fig. 5a with Supplementary Figure S8). In particular, the real part of $\Delta_X(\boldsymbol{k})$ is negligible; hence the $c$ and $v$ bands, renormalized by electron-hole pairing, each remain doubly degenerate.

# References

1. Keldysh, L. & Kopaev, Y. Possible instability of semimetallic state toward Coulomb interaction. *Sov. Phys. Sol. State* **6**, 2219 (1964).
2. Cloizeaux, J. D. Exciton instability and crystallographic anomalies in semiconductors. *Journal of Physics and Chemistry of Solids* **26**, 259–266 (1965).
3. Jerome, D., Rice, T. M. & Kohn, W. Excitonic Insulator. *Phys. Rev.* **158**, 462–475 (1967).
4. Halperin, B. I. & Rice, T. M. Possible Anomalies at a Semimetal-Semiconductor Transistion. *Rev. Mod. Phys.* **40**, 755–766 (1968).
5. Rohwer, T. *et al.* Collapse of long-range charge order tracked by time-resolved photoemission at high momenta. *Nature* **471**, 490–493 (2011).

6. Kogar, A. *et al.* Signatures of exciton condensation in a transition metal dichalcogenide. *Science* **358**, 1314–1317 (2017).
7. Lu, Y. F. *et al.* Zero-gap semiconductor to excitonic insulator transition in $Ta_2NiSe_5$. *Nature Communications* **8**, 14408 (2017).
8. Werdehausen, D. *et al.* Coherent order parameter oscillations in the ground state of the excitonic insulator $Ta_2NiSe_5$. *Science Advances* **4**, eaap8652 (2018).
9. Varsano, D. *et al.* Carbon nanotubes as excitonic insulators. *Nature Communications* **8**, 1461 (2017).
10. Ataei, S. S., Varsano, D., Molinari, E. & Rontani, M. Evidence of ideal excitonic insulator in bulk $MoS_2$ under pressure. *PNAS* **118**, (2021).
11. Eisenstein, J. P. & MacDonald, A. H. Bose–Einstein condensation of excitons in bilayer electron systems. *Nature* **432**, 691–694 (2004).
12. Nandi, D., Finck, A. D. K., Eisenstein, J. P., Pfeiffer, L. N. & West, K. W. Exciton condensation and perfect Coulomb drag. *Nature* **488**, 481–484 (2012).
13. Li, J. I. A., Taniguchi, T., Watanabe, K., Hone, J. & Dean, C. R. Excitonic superfluid phase in double bilayer graphene. *Nature Physics* **13**, 751–755 (2017).
14. Liu, X., Watanabe, K., Taniguchi, T., Halperin, B. I. & Kim, P. Quantum Hall drag of exciton condensate in graphene. *Nature Physics* **13**, 746–750 (2017).
15. Qian, X., Liu, J., Fu, L. & Li, J. Quantum spin Hall effect in two-dimensional transition metal dichalcogenides. *Science* **346**, 1344–1347 (2014).
16. Fei, Z. *et al.* Edge conduction in monolayer $WTe_2$. *Nature Physics* **13**, 677–682 (2017).
17. Wu, S. *et al.* Observation of the quantum spin Hall effect up to 100 kelvin in a monolayer crystal. *Science* **359**, 76–79 (2018).
18. Sajadi, E. *et al.* Gate-induced superconductivity in a monolayer topological insulator. *Science* **362**, 922–925 (2018).
19. Fatemi, V. *et al.* Electrically tunable low-density superconductivity in a monolayer topological insulator. *Science* **362**, 926–929 (2018).
20. Tang, S. *et al.* Quantum spin Hall state in monolayer 1T'-$WTe_2$. *Nature Physics* **13**, 683–687 (2017).
21. Cucchi, I. *et al.* Microfocus Laser–Angle-Resolved Photoemission on Encapsulated Mono-, Bi-, and Few-Layer 1T'-$WTe_2$. *Nano Lett.* **19**, 554–560 (2019).
22. Zheng, F. *et al.* On the Quantum Spin Hall Gap of Monolayer 1T'-$WTe_2$. *Advanced Materials* **28**, 4845–4851 (2016).
23. Ok, S. *et al.* Custodial glide symmetry of quantum spin Hall edge modes in monolayer $WTe_2$. *Phys. Rev. B* **99**, 121105 (2019).
24. Song, Y.-H. *et al.* Observation of Coulomb gap in the quantum spin Hall candidate single-layer 1T'-$WTe_2$. *Nature Communications* **9**, 4071 (2018).
25. Varsano, D., Palummo, M., Molinari, E. & Rontani, M. A monolayer transition-metal dichalcogenide as a topological excitonic insulator. *Nature Nanotechnology* **15**, 367–372 (2020).
26. Jia, Y. *et al.* Evidence for a Monolayer Excitonic Insulator. *arXiv:2010.05390 [cond-mat]* (2020).
27. Lee, P. A. Quantum oscillations in the activated conductivity in excitonic insulators: possible application to monolayer $WTe_2$. *arXiv:2010.09653 [cond-mat]* (2020).

28. Kwan, Y. H., Devakul, T., Sondhi, S. L. & Parameswaran, S. A. Theory of competing excitonic orders in insulating WTe$_2$ monolayers. *arXiv:2012.05255 [cond-mat]* (2020).
29. Zhao, W. *et al.* Determination of the spin axis in quantum spin Hall insulator monolayer WTe$_2$. *arXiv:2010.09986 [cond-mat]* (2021).
30. Wang, P. *et al.* Landau quantization and highly mobile fermions in an insulator. *Nature* **589**, 225–229 (2021).
31. Shi, Y. *et al.* Imaging quantum spin Hall edges in monolayer WTe$_2$. *Science Advances* **5**, eaat8799 (2019).
32. Tang, S. *et al.* Quantum spin Hall state in monolayer 1T'-WTe 2. *Nature Physics* **13**, 683–687 (2017).
33. Lupke, F. *et al.* Proximity-induced superconducting gap in the quantum spin Hall edge state of monolayer WTe$_2$. *Nature Physics* **16**, 526–530 (2020).
34. Zittartz, J. Anisotropy Effects in the Excitonic Insulator. *Phys. Rev.* **162**, 752–758 (1967).
35. Fei, Z. *et al.* Ferroelectric switching of a two-dimensional metal. *Nature* **560**, 336–339 (2018).
36. Giannozzi, P. *et al.* Advanced capabilities for materials modelling with Quantum ESPRESSO. *J. Phys.: Condens. Matter* **29**, 465901 (2017).
37. Hamann, D. R. Optimized norm-conserving Vanderbilt pseudopotentials. *Phys. Rev. B* **88**, 085117 (2013).
38. Onida, G., Reining, L. & Rubio, A. Electronic excitations: density-functional versus many-body Green's-function approaches. *Rev. Mod. Phys.* **74**, 601–659 (2002).
39. Hybertsen, M. S. & Louie, S. G. Electron correlation in semiconductors and insulators: Band gaps and quasiparticle energies. *Phys. Rev. B* **34**, 5390–5413 (1986).
40. Strinati, G. Application of the Green's functions method to the study of the optical properties of semiconductors. *Riv. Nuovo Cim.* **11**, 1–86 (1988).
41. Marini, A., Hogan, C., Grüning, M. & Varsano, D. Yambo: An ab initio tool for excited state calculations. *Computer Physics Communications* **180**, 1392–1403 (2009).
42. Sangalli, D. *et al.* Many-body perturbation theory calculations using the yambo code. *J. Phys.: Condens. Matter* **31**, 325902 (2019).
43. Elliott, J. D. *et al.* Surface susceptibility and conductivity of MoS$_2$ and WSe$_2$ monolayers: A first-principles and ellipsometry characterization. *Phys. Rev. B* **101**, 045414 (2020).
44. Rozzi, C. A., Varsano, D., Marini, A., Gross, E. K. U. & Rubio, A. Exact Coulomb cutoff technique for supercell calculations. *Phys. Rev. B* **73**, 205119 (2006).
45. Muechler, L., Alexandradinata, A., Neupert, T. & Car, R. Topological nonsymmorphic metals from band inversion. *Phys. Rev. X* **6**, 041069 (2016).
46. Monney, C. *et al.* Spontaneous exciton condensation in 1T-TiSe$_2$: BCS-like approach. *Phys. Rev. B* **79**, 045116 (2009).

**Code availability statement.** Many-body perturbation theory calculations were performed by means of the codes Yambo (http://www.yambo-code.org/) and Quantum ESPRESSO (http://www.quantum-espresso.org), which are both open source software. Results for the two-band model were obtained through custom Fortran codes that are available from M.R. upon reasonable request.

**Acknowledgements.** The experiments and analysis were supported primarily by NSF DMR awards MRSEC 1719797 and EAGER 1936697. Materials synthesis at UW was partially supported by the Gordon and Betty Moore Foundation's EPiQS Initiative, Grant GBMF6759 to JHC. X.H. and Y.-T.C. acknowledge support from the NSF under award no. DMR-2004701 and a Hellman Fellowship award. On the theory side, S.S.A., D.V., E.M., and M.R. were supported in part by the MaX European Centre of Excellence ("MAterials design at the eXascale", www.max-centre.eu) funded by the European Union H2020-INFRAEDI-2018-1 programme, grant No. 824143. D.V., E.M., and M.R. were also supported by the Italian national program PRIN2017 No. 2017BZPKSZ 'Excitonic insulator in two-dimensional long-range interacting systems (EXC-INS)'. M.P. was supported by INFN Time2quest project and by the European Union H2020-MSCA-RISE project DiSeTCom, grant No. 823728. S.S.A., M.P., D.V., E.M., and M.R. acknowledge access to the Marconi supercomputing system based at CINECA, Italy, through PRACE and ISCRA programs.

# Supplementary Information:
# Evidence for equilibrium excitons and exciton condensation in monolayer WTe$_2$

Bosong Sun[1], Wenjin Zhao[1], Tauno Palomaki[1], Zaiyao Fei[1], Elliott Runburg[1], Paul Malinowski[1], Xiong Huang[2,3], John Cenker[1], Yong-Tao Cui[2], Jiun-Haw Chu[1], Xiaodong Xu[1], S. Samaneh Ataei[4,5], Daniele Varsano[4], Maurizia Palummo[6], Elisa Molinari[4,7], Massimo Rontani[4*], David H. Cobden[1*]

[1]*Department of Physics, University of Washington, Seattle, WA 98195, USA*
[2]*Department of Physics and Astronomy, University of California, Riverside, CA 92521, USA*
[3]*Department of Materials Science and Engineering, University of California, Riverside, CA 92521, USA*
[4]*CNR-NANO, Via Campi 213a, 41125 Modena, Italy*
[5]*Dept of Physics, Shahid Beheshti University, Evin, Tehran 1983969411, Iran*
[6]*INFN, Dept of Physics, University of Rome Tor Vergata, Via della Ricerca Scientifica 1, 00133 Roma, Italy*
[7]*Dept of Physics, Informatics and Mathematics, University of Modena and Reggio Emilia, Via Campi 213a, 41125 Modena, Italy*
Email: massimo.rontani@nano.cnr.it; cobden@uw.edu

## Contents

## SI-1 Devices

We present measurements on five monolayer WTe$_2$ devices with somewhat different configurations. Devices MW2 and MW3 have top graphite gate and bottom SiO$_2$/silicon substrate gate. Device MW10 has bottom graphite gate and no top graphite gate, to allow MIM measurements to be performed. Devices MW12 and MW15 have an additional graphene layer with multiple contacts in parallel with the WTe$_2$ allowing chemical potential measurements. We will also show measurements on bilayer devices BW4 and B2.

The basic fabrication process of a typical device is essentially as follows. First, graphite and hBN crystals are mechanically exfoliated onto a SiO$_2$/Si substrate. Using the van der Waals (vdW) transfer technique with polycarbonate/polydimethylsiloxane (PC/PDMS) "stamp"[1], the few-layer graphite bottom gate is covered by an hBN flake (bottom hBN). After dissolving the PC polymer in chloroform, the hBN/graphite is annealed at 400 °C for 2h for cleaning. Next, Pt metal contacts (~7 nm) are patterned on the hBN by e-beam lithography and lift-off. Then, WTe$_2$ crystals are exfoliated in a nitrogen-filled glovebox (O$_2$ and H$_2$O concentrations below 0.5 ppm) and monolayers or bilayers are optically identified. A stack of top-gate graphite on hBN, or just hBN, is moved into the glovebox on a stamp, the WTe$_2$ is picked up under it, and the result is put down on the Pt contacts on hBN. After the PC polymer is dissolved, another step of e-beam lithography, metallization (~7 nm V, ~70 nm Au) and lift-off is used to define wire-bonding pads and connections to the Pt contacts and the top/bottom graphite gates. For devices MW12, MW15 and B2, an extra round of e-beam lithography/metal film deposition/vdW transfer was performed to make the additional graphene layer with its own contacts. Steps in the fabrication of MW15 are shown in Fig. S1. The device capacitance parameters are given in tables T1 and T2. These are purely geometric capacitances, but corrections for finite density of states (quantum capacitance effects) are at the percent level in all cases. This can be deduced from the measurements shown in Fig. S4: the total measured variation of the chemical potential $\mu$ in the WTe$_2$ (~60 meV) is two orders of magnitude smaller than the applied gate voltage (~5 V).

| Device label | WTe$_2$ | Top hBN (nm) | Bottom hBN (nm) | $C_{tg}$ (10$^{12}$e/cm$^2$V) | $C_{bg}$ (10$^{12}$e/cm$^2$V) |
|---|---|---|---|---|---|
| MW2 | Monolayer | 9.2 | 17.5 | 2.4 | Si gate |
| MW3 | Monolayer | 11.4 | 14 | 1.94 | Si gate |
| MW10 | Monolayer | N/A | 18 | N/A | 0.92 |
| BW4 | Bilayer | 10 | 20.7 | 2.2 | 1.07 |

**Table T1.** Thickness of the top and bottom hBN and corresponding areal geometric capacitances $C_{tg}$ and $C_{bg}$ between adjacent conductors. The gate-induced electron density is $n_g = C_g V_g / e$, where $C_g = \epsilon_{hBN} \epsilon_0 / d_{hBN}$. The effective dielectric constant of hBN is known to depend somewhat on the hBN thickness[2,3]. We used this way for some adjustment to make the critical density $n_{ce}$ for the insulator-metal transition consistent between devices, requiring $\epsilon_{hBN} = 4$ in devices MW2, MW3, BW4 and $\epsilon_{hBN} = 3$ in device MW10.

| Device | WTe$_2$ | Top hBN (nm) | Middle hBN (nm) | Bottom hBN (nm) | $C_{tg}$ (10$^{12}$e/cm$^2$V) | $C_{mg}$ (10$^{12}$e/cm$^2$V) | $C_{bg}$ (10$^{12}$e/cm$^2$V) |
|---|---|---|---|---|---|---|---|
| MW12 | Monolayer | 15 | 9 | N/A | 1.1 | 1.84 | N/A |
| MW15 | Monolayer | 27 | 20 | 11 | 0.61 | 0.83 | 1.51 |
| B2 | Bilayer | 25 | 8 | N/A | 0.89 | 2.76 | N/A |

**Table T2.** Thickness of the top/middle/bottom hBN and corresponding areal geometric capacitances $C_{tg}$, $C_{mg}$, $C_{bg}$ for devices including a graphene layer for chemical potential measurements, where $C_g = \epsilon_{hBN} \epsilon_0 / d_{hBN}$ and $C_{mg}$ is between the graphene and the WTe$_2$.

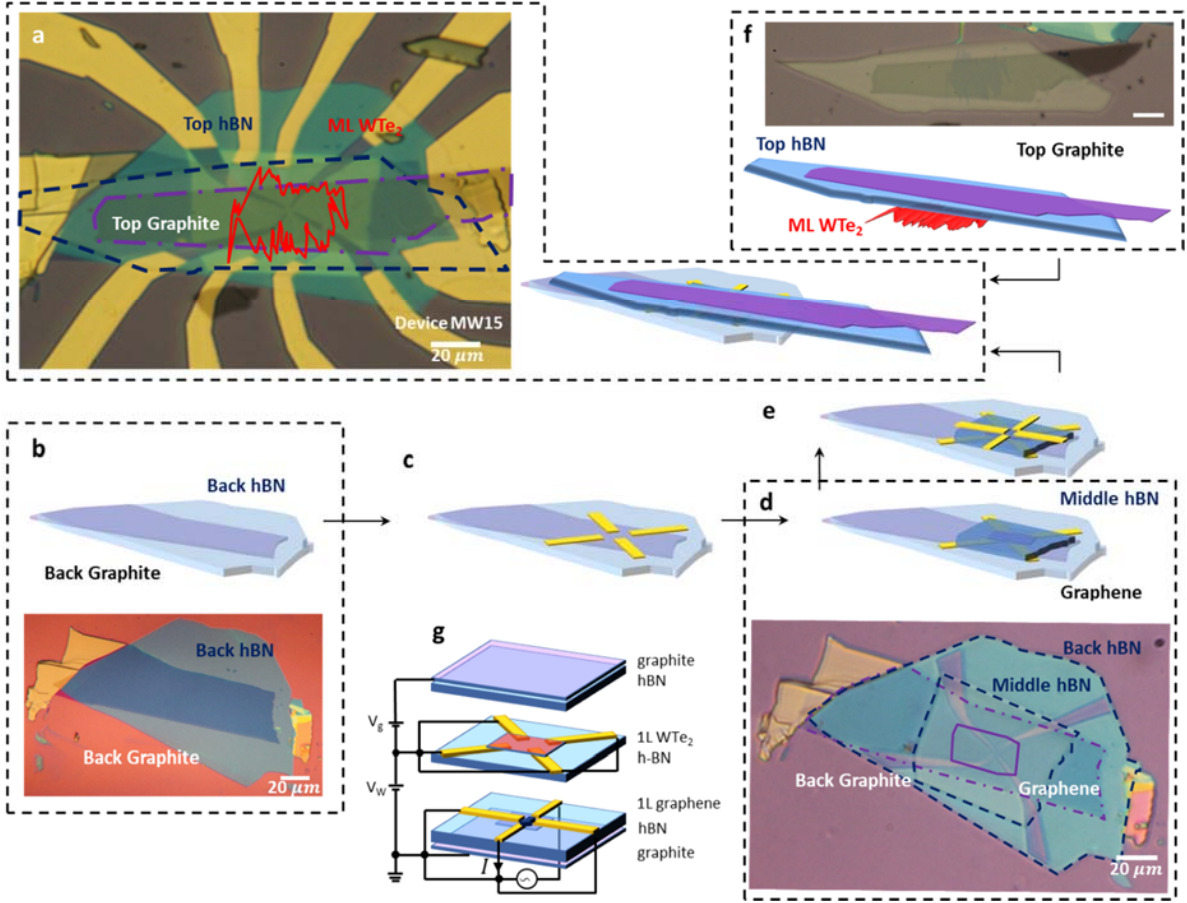

**Figure S1. Example of sample fabrication and measurement setup: device MW15.** (a) Optical image of device MW15. The monolayer WTe$_2$ is outlined in red. (b) Bottom hBN and graphite stack preparation. (c) Patterning of metal contacts (7 nm Pt) electrodes for graphene on bottom layer hBN. (d) Transfer of the middle layer hBN and graphene; optical image shows the transfer in progress. (e) Deposition of metal contacts electrodes (7 nm Pt) for monolayer WTe$_2$ on middle layer hBN. (f) Picking up top graphite/top hBN/monolayer (1L) WTe$_2$: the optical image shows the pick-up in progress. After transferring the top stack to the prefabricated bottom part, we end up with the device MW15 shown in (a). (g) Setup used with this device to measure $\mu$ vs $n_e$ (see Fig. 3a in main text). We use a "current-focusing" geometry employing four electrodes in each layer, as shown. The idea is as follows: when we use the graphene as detector, we apply a bias to one contact with the other three at ground potential, and measure the current to ground in the opposite contact. This current only flows through the center of the cross and so depends mostly on the conductivity of the graphene in that region, which in turn is sensitive to the electric field conditions of only the WTe$_2$ directly above it. This helps exclude effects of edges, cracks and nonuniformity in the WTe$_2$ by probing only a small central region of it (see also SI-5 & SI-6).

**SI-2 Microwave impedance microscopy (MIM)**

Scanning microwave impedance microscopy (MIM) measurements (shown in Fig. 2, main text) were performed in a home-built cryogenic scanning probe microscope[4]. A small microwave signal of ~0.1 $\mu$W at a fixed frequency in the range of 1-10 GHz was delivered to a chemically etched tungsten tip. The reflected signal was analyzed to extract the demodulated output channels, MIM-Im and MIM-Re, which are proportional to the imaginary and real parts of the admittance between the tip and the sample, respectively. The MIM-Im characterizes the sample's capability to screen the oscillating electric field at the tip, while the MIM-Re characterizes the amount of dissipation generated by the screening current. To enhance the MIM signal quality, the tip was excited to oscillate at a frequency of ~32 kHz with an amplitude of ~8 nm. The resulting oscillation amplitudes of MIM-Im and MIM-Re were then extracted using a lock-in amplifier to yield $d(\text{MIM-Im})/dz$ and $d(\text{MIM-Re})/dz$, respectively. The $d(\text{MIM})/dz$ signals are free of fluctuating backgrounds, thus enabling more quantitative analysis. To extract quantitative values of local conductivity from MIM measurements, finite element simulation was performed using COSMOL with known geometrical parameters of the sample. The imaginary and real parts of the tip-sample admittance $Y_{ts}$ were calculated as a function of the sample resistivity at different tip-sample separations, from which $dY_{ts}/dz$ were extracted. MIM signals measured over regions with known electrical properties were used to scale the simulated $dY_{ts}/dz$ to the measured $d(\text{MIM})/dz$. These reference regions include the graphite contact which corresponds to the highly conductive limit, and the nearby hBN substrate with no sample or graphite contact which corresponds to the highly insulating limit.

## SI-3 Raman spectroscopy measurements

To perform Raman, a sample was excited at normal incidence by a 632.8 nm excitation laser in backscattering geometry. The laser power was kept below 400 $\mu$W to prevent sample degradation. The scattered light was collected and dispersed by a 1200 mm$^{-1}$ groove density grating and detected by a cooled charge-coupled device (CCD) with an integration time of 5 minutes. BragGrate™ notch filters were used to reject Rayleigh scatter down to 5 cm$^{-1}$. A linear polarizer and half-wave plate (HWP) placed between the notch filter and the sample allow the detection of Raman features that are co-linear with the excitation laser. For angle dependence, the HWP is continuously rotated by 5° continuously until we scan through a full 360°. To save time, we only scanned from 0° to 180° in the temperature dependent measurements and replicated that data for polarizations from 180° to 360°.

***Determination of crystal axes.*** Before assembling a device, the $x$-axis (W-chain axis) of the monolayer WTe$_2$ could be guessed from the orientation of nearby larger flakes and tape residue. In the completed devices, polarization-resolved Raman spectroscopy was then used to confirm the alignment. The Raman spectrum of a WTe$_2$ monolayer is shown in Fig. S2. The peaks near 160 cm$^{-1}$ and 210 cm$^{-1}$ are labelled P10 and P11. In the colinear configuration described above, both are strongest along the $y$-axis in the monolayer, as seen in Fig. S2b, and bulk samples[5]. We therefore used several bulk WTe$_2$ flakes with easily identifiable crystal axes to calibrate and verify the alignment between the optical image and incident light polarization.

***Search for charge density waves (CDWs).*** Normally, distinct changes in the Raman spectrum of a layered sample provide a signature of CDW formation due to the associated symmetry change (see, for example, Ref. 6 for the case of 1T-TiSe$_2$). Our detailed study of the temperature dependence of the spectrum of 1L WTe$_2$ from room temperature down to 15 K (Fig. S2c), well below the appearance of the insulating state, reveals no indication of any such changes. This is consistent with the fact that no clear signs of CDWs have been observed in STM (see, e.g. Refs. 7–9).

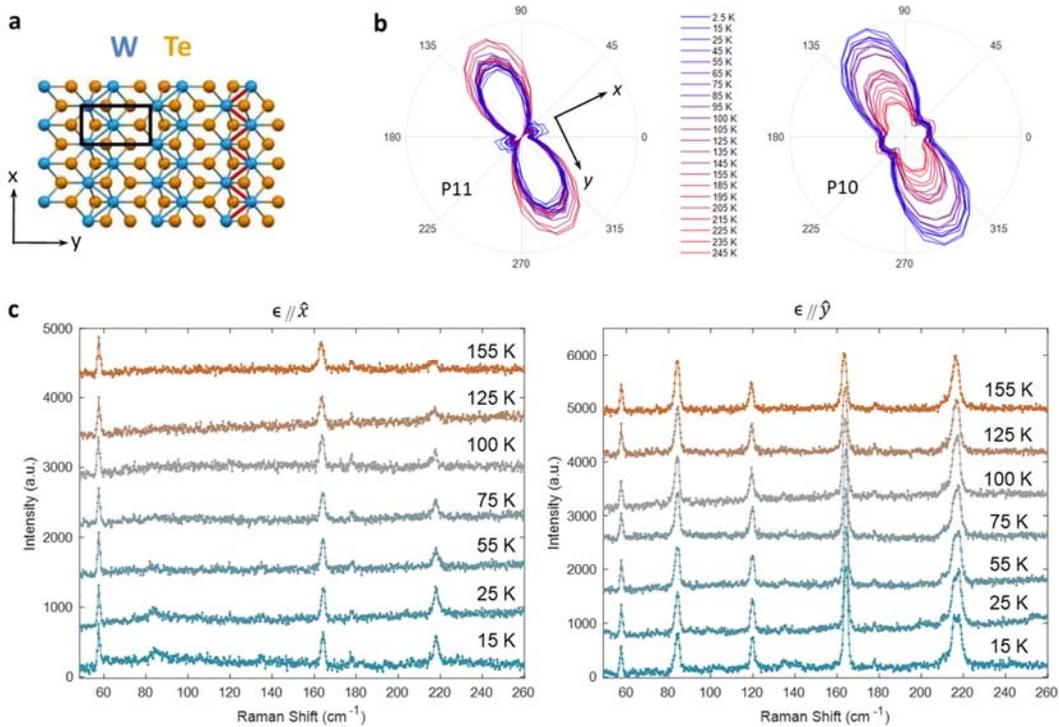

**Figure S2. Raman Spectroscopy.** (a) Structure of 1L WTe$_2$ seen from above; the $x$-axis is taken along the W chains. (b) Anisotropy of peaks P11 (210 cm$^{-1}$) and P10 (160 cm$^{-1}$) for colinear polarization at a series of temperature. (c) Raman intensity for polarization parallel to $x$-axis and $y$-axis at a series of temperatures, showing no distinct changes across the entire temperature range that could indicate CDW formation.

## SI-4 Notes on anisotropy measurements

We here briefly provide more discussion of the measurements shown in Fig. 2c of the main text. Since the edges of monolayer WTe$_2$ conduct, and there are often cracks present which effectively produce internal edges, we use the "empty cross" contact pattern sketched in Fig. S3a to minimize the effect of such cracks and edges. In devices with no top gate, such as MW10, MIM can be used to identify empty crosses where there are no cracks nearby. We then apply a bias to one contact and collect the current to ground from the opposite contact with the other two contacts directly grounded. Most of the collected current then passes only through the center region of the cross and flows roughly parallel to the line joining the contacts, hence probing the conductivity in this direction. We confirmed this by simulating the current flow for a uniform conducting sheet with contacts in this shape using finite-element analysis algorithms, corresponding to solving an anisotropic Laplace equation with mixed boundary conditions. Fig. S3b shows the region probed by 95% of current flowing in this particular geometry, here taking the conductivity to be isotropic.

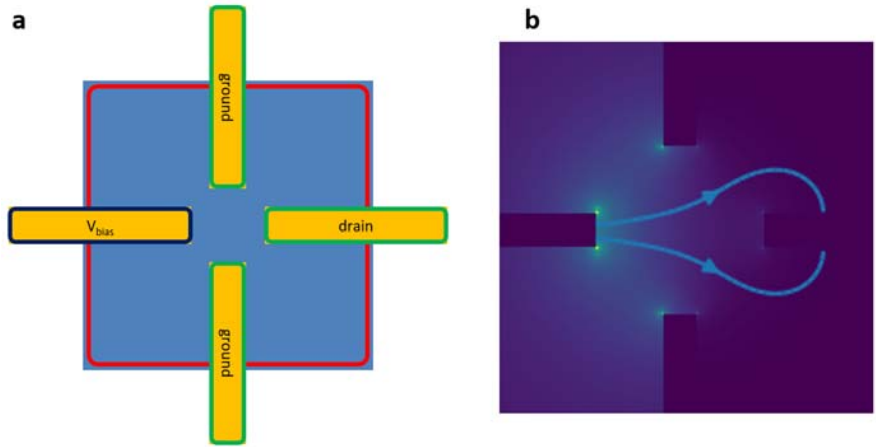

**Figure S3**. (a) Contact and boundary geometry for a current distribution simulation. Red line: boundary of 1L WTe$_2$ (assumed to be a uniform 2D conductor), where normal current density vanishes (and we neglect edge conduction). Blue line: equipotential at $V = V_{bias}$. Green line: equipotential at $V = 0$. (b) Current streamlines enclosing 95% of the current flowing from the left to the right contact.

**SI-5 Comparison of chemical potential in two devices. Signs of excitons at 300 K**

The measurements on device MW12, presented in Fig. 3 of the main text, are compared with similar measurements on similar device MW15 in Fig. S4. Their behavior is consistent, except that in MW15 the step in $\mu$ is not as steep and the "V" in the conductance is more rounded. It is very likely that this is a result of greater sample inhomogeneity in MW15. The measurements on MW15, however, extend to 300 K and show that some residual step is present in the chemical potential at the neutral point even at 300 K, even though the step height is much less than $kT$ (green bar), suggesting that excitonic effects are relevant up to room temperature.

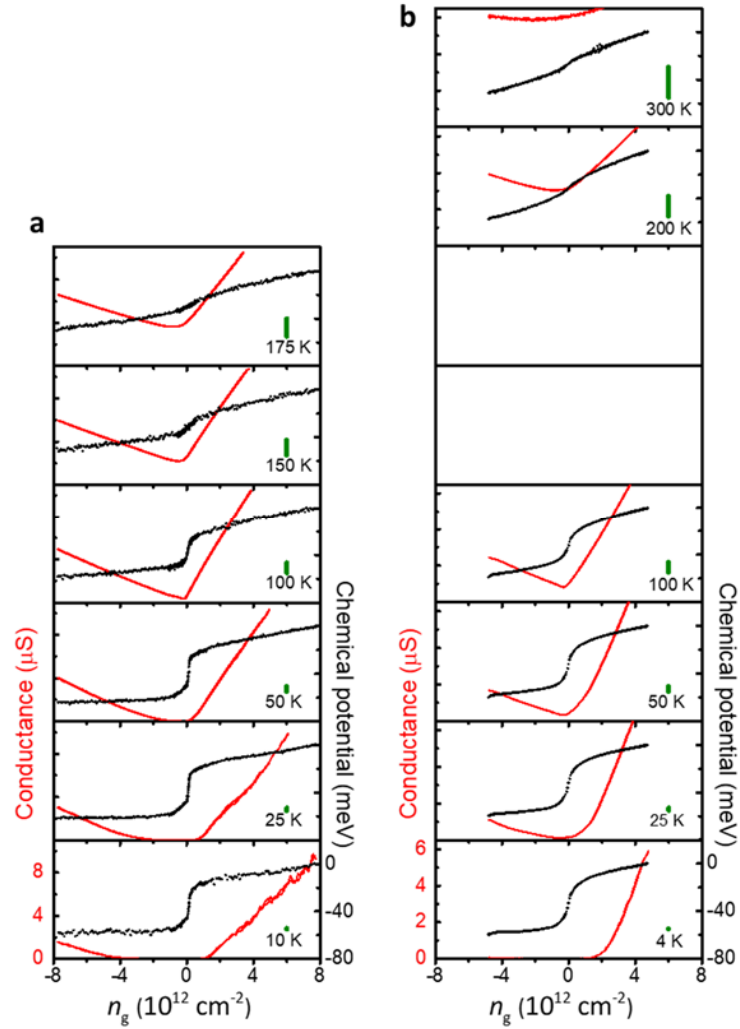

**Figure S4.** (a) Measurements of chemical potential $\mu$ (black) and conductance $G_p$ (red) vs. gate doping $n_g$ on 1L WTe$_2$ device MW12, reproduced from Fig 3(b) in the main text. (b) Comparable measurements on device MW15. The zero of $\mu$ is chosen arbitrarily at each temperature.

## SI-6 Notes on modeling the "V": mobility, overlapping bands, and chemical equilibrium

To aid the following discussion, Fig. 4a-c of the main text is reproduced as Fig. S5a-c.

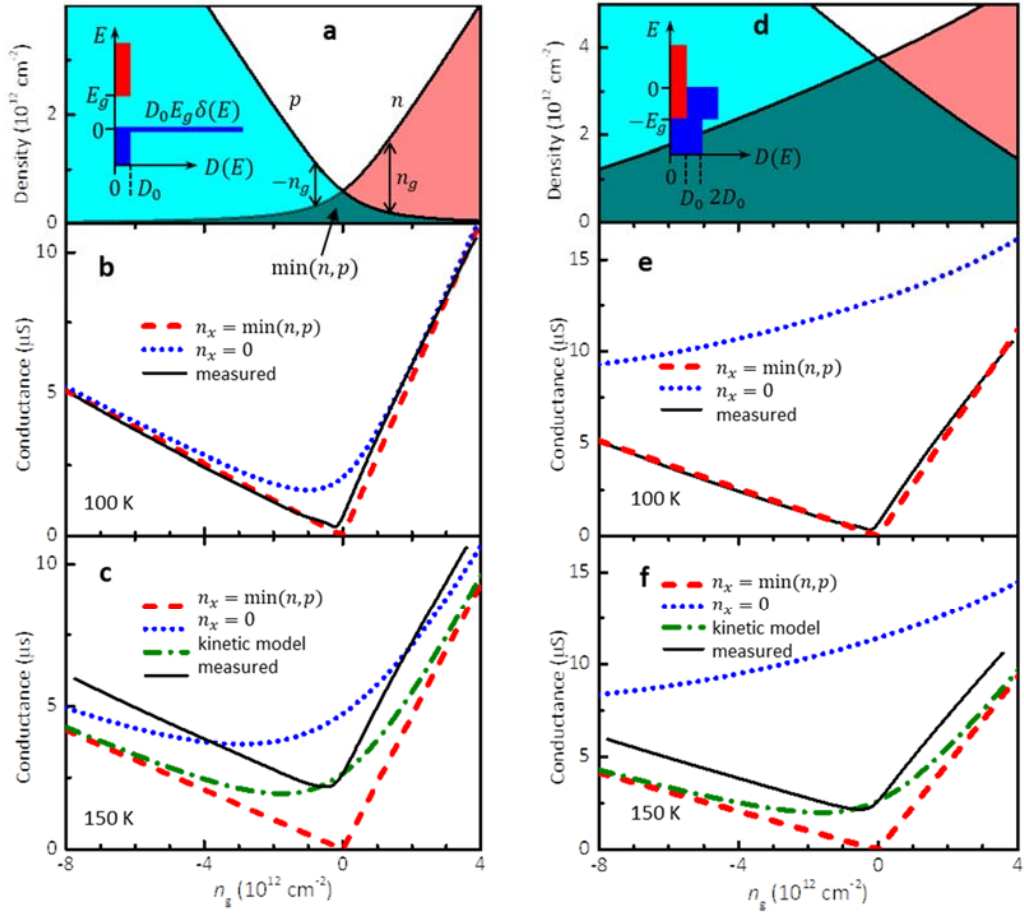

**Figure S5.** (a-c) Same as Fig. 4 in the main text, using the gapped density of states shown in Fig. 3 and in the inset to (a). (d-f) Corresponding graphs using overlapped bands, as shown in the inset to (d).

*Mobility.* In our argument that electron-hole pairing helps explain the sharp "V", we compared the conductance data at 100 K (black trace in Fig. S5b) with the simple model where all minority carriers are bound, $n_x = \min(n,p)$ (red dashed trace). Then, (neglecting contact resistance $R_c$ because $G_p$ is small near zero doping), $G_p = \beta^{-1}\sigma = \beta^{-1}\mu_e e n_g$ for $n_g > 0$ and $G_p = -\beta^{-1}\mu_h e n_g$ for $n_g < 0$. Since the geometric factor $\beta$ is unknown, we cannot determine the absolute mobilities accurately. In addition, we know the conductivity is anisotropic but we do not know the orientation of the current relative to the axis in this device. This is unimportant, however, as our objective is simply understanding the linear "V" shape. We therefore treat $\beta^{-1}e\mu_e$ and $\beta^{-1}e\mu_h$ as parameters to achieve the best fit at each temperature; their values are unimportant but for completeness at 100 K they are $\beta^{-1}e\mu_e = 2.72 \times 10^{-12}$ µS cm² and $\beta^{-1}e\mu_h = 0.64 \times 10^{-12}$ µS cm² and at 150 K they are $\beta^{-1}e\mu_e = 2.33 \times 10^{-12}$ µS cm² and $\beta^{-1}e\mu_h = 0.52 \times 10^{-12}$ µS cm².

*Effect of overlapping bands.* The blue dotted lines in Figs. S5b and c are plots of $\sigma = \mu_e n + \mu_h p$, where $n = \int_{-\infty}^{+\infty} D_c(E)f(E)dE$ and $p = \int_{-\infty}^{+\infty} D_v(E)(1-f(E))dE$ calculated using the gapped single-particle spectrum shown in the left inset taking $E_g = +43$ meV. To illustrate the effect of modifying $D(E)$, in Figs. S5d-f we show the result of doing the same calculations but with the $c$ and $v$ bands overlapping, i.e., with a negative gap $E_g = -40$ meV, as shown in the right inset. Here we have no $\delta$-function at the $v$ band edge and set $D_v/D_c = 1/2$ simply because $n_{ce}/n_{cp} \sim 2$. Although $n$ and $p$ are now much larger (Fig. S5d), and without excitons the model conductance varies smoothly with $n_g$ (Fig. S5e) at odds with the data, if we assume again that $n_x = \min(n,p)$ then naturally we get the same "V" shape, matching the data at 100 K (red dashed), as when using the gapped spectrum.

*Chemical equilibrium considerations.* As mentioned in the main text, one may also ask whether the observed behavior above 100 K could be explained by a variation of the exciton density with $n_g$. To make it clear that this is not feasible, we consider a variation based on simple chemical equilibrium between the excitons and free particles, i.e.,

$$n_x = K(n - n_x)(p - n_x),$$

where $K$ is an equilibrium constant[10]. For simplicity we assume $K$ is independent of $n_g$ for a given temperature. Solving for $n_x$ gives

$$n_x = \frac{1}{2}\left[n + p + \frac{1}{K} - \sqrt{\left(n + p + \frac{1}{K}\right)^2 - 4np}\right].$$

The green dash-dotted lines in Figs. S5c and S5f are plots of $\sigma = \mu_e(n - n_x) + \mu_h(p - n_x)$ using the above expression for $n_x$ with $K$ chosen to match the measured conductance at $n_g = 0$ and using the same $\beta^{-1}\mu_{e,h}$ values as before. The agreement with the data remains poor, irrespective of the single-particle spectrum. The chemical equilibrium condition does not prevent thermal smearing, and also it causes to $n_x$ to approach $\min(n,p)$ as $|n_g|$ increases, making the calculated conductance too small at high gate voltages.

## SI-7 Measurements on bilayer WTe$_2$

Conductance and chemical potential measurements on bilayer WTe$_2$ (Fig. S6) show striking similarities to those on monolayers. The conductance of bilayer is more straightforward to measure because it lacks edge conduction; on the other hand, the ferroelectric switching complicates the chemical potential measurements[11]. A sharp "V" develops in conductance vs gate voltage (Fig. S6a) at the neutral point at around 20 K, while there is a step in the chemical potential (Fig. S6b) of size ~10 meV that is fully developed below 20 K. This suggests that similar excitonic physics is at play in the bilayer but at a ~5 times smaller energy scale than that in the monolayer. Trilayer WTe$_2$ is metallic.

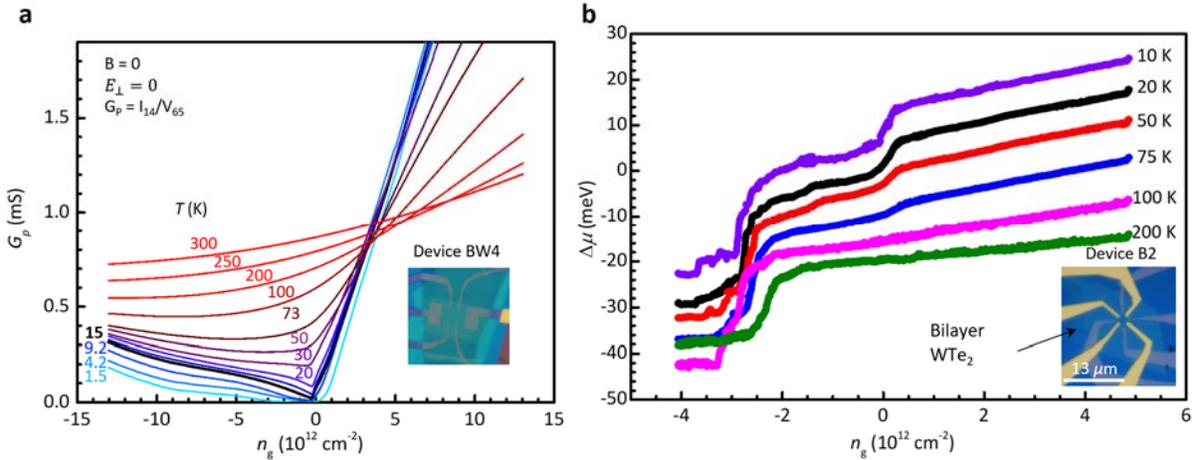

**Figure S6** (a) Conductance characteristics of 2L WTe$_2$ device BW4 (for zero displacement field). Here the inverse 4-terminal resistance is plotted. (b) Chemical potential. Ferroelectric switching in the bilayer limits the range of the chemical potential measurement: these measurements are all performed sweeping $n_g$ downwards in a single polarization state.

## SI-8 Scaling of excitation energy of lowest direct exciton with inverse number of $k$ points, $n_k$

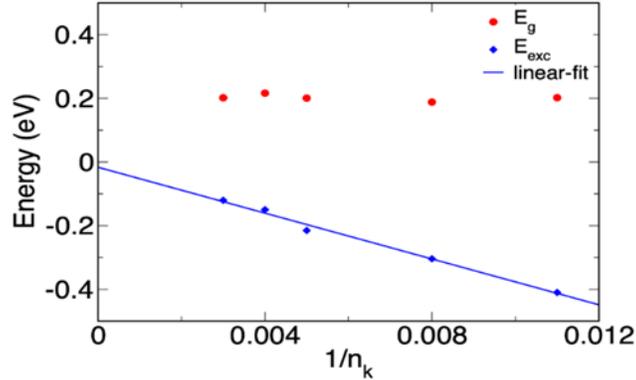

**Figure S7.** Plot of the direct band gap ($E_g$, red dots; not to be confused with the absolute, indirect band gap) and excitation energy of the lowest exciton with momentum $q = 0$ (blue diamonds, $E_{exc}$), versus inverse number of irreducible $k$ points, $1/n_k$. The solid line is the linear regression of the excitation energy on $1/n_k$. The difference between the energy extrapolated for $1/n_k = 0$ and the value obtained for the largest implemented sampling is 100 meV, which is the uncertainty induced by the numerical discretization of $k$-space, shown by the error bars in Fig. 5b (main text).

## SI-9 Band dispersion perpendicular to ΓΛ cut of the Brillouin zone

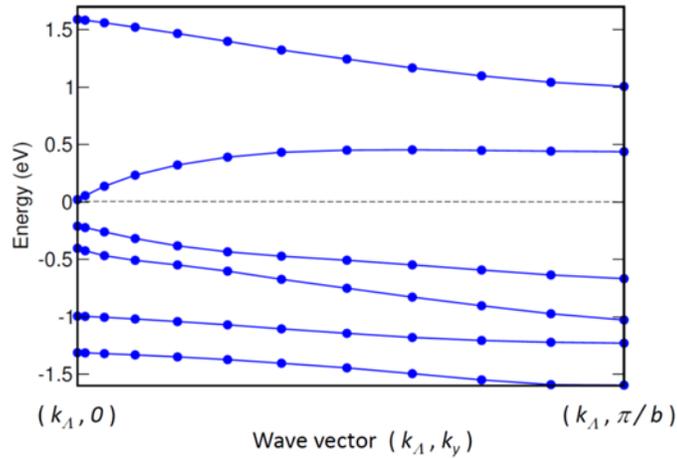

**Figure S8.** Calculated bands from first principles (DFT-PBE0 level, see Methods) along the cut of the Brillouin zone parallel to ΓY and intersecting the Λ point. The plot is a companion to Fig. 5a in the main text which is along the ΓX cut.

## SI-10 Crystal growth

WTe$_2$ single crystals were grown out of a Te-rich self-flux with the following procedure[12]. Elemental W power and Te shot were loaded into alumina crucibles in the molar ratio 1:50 with a total charge of 3.5 g. The crucibles were loaded into quartz tubes and sealed under an evacuated argon atmosphere. The tubes were then placed in a box furnace and heated to 1000 °C over 12 hours, held at 1000 °C for 12 hours, cooled for 460 °C over 100 hours, and finally decanted at 460 °C via centrifuge. This procedure resulted in large, high quality single crystals of WTe$_2$ with ribbon-like morphology and typical dimensions 5 mm × 0.3 mm × 0.01 mm.

**References**
1. Zomer, P. J., Guimarães, M. H. D., Brant, J. C., Tombros, N. & van Wees, B. J. Fast pick up technique for high quality heterostructures of bilayer graphene and hexagonal boron nitride. *Appl. Phys. Lett.* **105**, 013101 (2014).
2. Zhao, W. *et al.* Magnetic proximity and nonreciprocal current switching in a monolayer $WTe_2$ helical edge. *Nature Materials* **19**, 503–507 (2020).
3. Xu, S.-Y. *et al.* Electrically switchable Berry curvature dipole in the monolayer topological insulator $WTe_2$. *Nature Physics* **14**, 900–906 (2018).
4. Shi, Y. *et al.* Imaging quantum spin Hall edges in monolayer $WTe_2$. *Science Advances* **5**, eaat8799 (2019).
5. Kim, M. *et al.* Determination of the thickness and orientation of few-layer tungsten ditelluride using polarized Raman spectroscopy. *2D Mater.* **3**, 034004 (2016).
6. Duong, D. L. *et al.* Raman Characterization of the Charge Density Wave Phase of $1T-TiSe_2$: From Bulk to Atomically Thin Layers. *ACS Nano* **11**, 1034–1040 (2017).
7. Tang, S. *et al.* Quantum spin Hall state in monolayer 1T'-$WTe_2$. *Nature Physics* **13**, 683–687 (2017).
8. Song, Y.-H. *et al.* Observation of Coulomb gap in the quantum spin Hall candidate single-layer 1T'-$WTe_2$. *Nature Communications* **9**, 4071 (2018).
9. Lupke, F. *et al.* Proximity-induced superconducting gap in the quantum spin Hall edge state of monolayer $WTe_2$. *Nature Physics* **16**, 526–530 (2020).
10. Siviniant, J., Scalbert, D., Kavokin, A. V., Coquillat, D. & Lascaray, J.-P. Chemical equilibrium between excitons, electrons, and negatively charged excitons in semiconductor quantum wells. *Phys. Rev. B* **59**, 1602–1604 (1999).
11. Fei, Z. *et al.* Ferroelectric switching of a two-dimensional metal. *Nature* **560**, 336–339 (2018).
12. Wu, Y. *et al.* Temperature-Induced Lifshitz Transition in $WTe_2$. *Phys. Rev. Lett.* **115**, 166602 (2015).